\documentclass[]{spie}  

\usepackage{graphicx}
\usepackage{amsmath}
\usepackage{amssymb}

\newcommand\ud{\mathrm{d}}

\newcommand\Fermi{\mathrm{F}}
\newcommand\eff{\mathrm{eff}}

\def\bs{\boldsymbol}

\newcommand{\maxxi}[1]{{\raise-2.mm\hbox{$\textstyle  {\rm max}
 \atop {\raise1.7mm\hbox{$\scriptscriptstyle \xi$}}$}}\! \left\{ #1\right\}}

\def\toinfty{\hbox{\space \raise-1.5mm\hbox{ $\textstyle  \longrightarrow
 \atop {\raise1mm\hbox{$\scriptstyle \r\to\infty$}}$} \space}}

\def\C{\mathrm{k}}

\title{Modeling of optical, transport, and thermodynamic properties of Al metal irradiated by intense femtosecond laser~pulses}

\author{
Konstantin V. Khishchenko, 
Mikhail E. Veysman, 
Nikolay E. Andreev, \\ 
Vladimir E. Fortov,
Pavel R. Levashov, 
and Mikhail E. Povarnitsyn 
\skiplinehalf 
Joint Institute for High Temperatures, Russian Academy of Sciences, \\ Izhorskaya Street~13 Bldg~2, Moscow 125412, Russia}

\authorinfo{Further author information: (Send correspondence to K.V.K.)\\
K.V.K.: E-mail: konst@ihed.ras.ru, Telephone: +7 495 484 24 56 
}

\begin{document}
  \maketitle


\begin{abstract}
A theoretical model is developed for the interaction of intense femtosecond laser pulses with solid targets on the basis of the two-temperature equation of state for an irradiated substance. It allows the description of the dynamics of the plasma formation and expansion. Comparison of available experimental data on the amplitude and phase of the complex reflection coefficient of aluminum with the simulation results provides new information on the transport coefficients and absorption capacity of the strongly coupled Al plasma over a wide range of temperatures and pressures. 
\end{abstract}




\section{Introduction}

The optical properties of a plasma formed on the surface of an aluminum target subjected to ultrashort laser pulses are investigated in interferometric experiments~\cite{ Gauthier_Geindre_Audebert_Bastiani_Quoix_Grillon_Mysyrowicz_Antonetti_Mancini_1997, Grimes_Rundquist_Lee_Downer_1999, Widmann_Guethlein_Foord_Cauble_Patterson_Price_Rogers_Springer_Stewart_Ng_Ao_Forsman_2001, Agranat_Andreev_Ashitkov_Veisman_Levashov_Ovchinnikov_Sitnikov_Fortov_Khishchenko_2007 } at irradiation intensities up to $I \sim 10^{16}$~W/cm$^2$. In the present work, we analyzed data~\cite{ Agranat_Andreev_Ashitkov_Veisman_Levashov_Ovchinnikov_Sitnikov_Fortov_Khishchenko_2007} recently obtained with femtosecond pulses of $I \lesssim 10^{14}$~W/cm$^2$. 
Such laser pulses produce a thin layer of the strongly coupled plasma with an electron temperature up to $T \sim 10$~eV on the initially cold metal target. We developed semiempirical models of optical, transport, and thermodynamic properties of Al over the whole range of temperatures realized in a numerical simulation of the laser--metal interaction processes. In contrast to previous simulations~\cite{ Gauthier_Geindre_Audebert_Bastiani_Quoix_Grillon_Mysyrowicz_Antonetti_Mancini_1997, Grimes_Rundquist_Lee_Downer_1999, Velichko_Urlin_Yakutov_2000, Widmann_Guethlein_Foord_Cauble_Patterson_Price_Rogers_Springer_Stewart_Ng_Ao_Forsman_2001}, particular attention is focused on the initial stage ($t \leqslant 1$~ps) of the heating and expansion of the plasma under the conditions of the undeveloped hydrodynamic motion of ions. 

Comparison of the numerical simulation results with experimental data on the complex reflection coefficient~\cite{ Agranat_Andreev_Ashitkov_Veisman_Levashov_Ovchinnikov_Sitnikov_Fortov_Khishchenko_2007} makes it possible to determine the free parameters in the formulated expressions for the effective electron collision frequency, in contrast to the models discussed elsewhere~\cite{ Velichko_Urlin_Yakutov_2000, Eidmann_Meyer-ter-Vehn_Schlegel_Huller_2000, Colombier_Combis_Bonneau_Le_Harzic_Audouard_2005 }. In contrast to the approach proposed in Ref.~\citenum{Isakov_Kanavin_Uryupin_2006_10}, this study is free of the assumptions that the metal is not melted and the electron temperature is much lower than the Fermi energy. This allows us to use the developed models in a broad region of states of the strongly coupled plasma formed on the target surface. 

The detailed description of the experimental technique for simultaneous measurements of both the amplitude and phase of the complex reflection coefficient can be found in Ref.~\citenum{ Agranat_Andreev_Ashitkov_Veisman_Levashov_Ovchinnikov_Sitnikov_Fortov_Khishchenko_2007}. The source of radiation in the experiments of interest~\cite{ Agranat_Andreev_Ashitkov_Veisman_Levashov_Ovchinnikov_Sitnikov_Fortov_Khishchenko_2007} was a Cr:forsterite laser system generating femtosecond pulses at a wavelength of $\lambda_1 = 1240$~nm~\cite{ Agranat_Ashitkov_Ivanov_Konyashchenko_Ovchinnikov_Fortov_2004_06}. The full width at half maximum (FWHM) of the pulses that was measured using an autocorrelator of the non-collinear second harmonic was equal to $\tau_L \simeq 110$~fs in the experiments for the sech$^2$ envelope shape. The time profile of the pulse was measured in a wide range of power using the third-harmonic correlator. The ratio (contrast) of the intensity at the pulse maximum to the intensity 1 and 2~ps before the maximum was no less than $10^4$ and $10^6$ respectively~\cite{ Agranat_Andreev_Ashitkov_Ovchinnikov_Sitnikov_Fortov_Shevelko_2006}. 

For the discussed below experimental data~\cite{ Agranat_Andreev_Ashitkov_Veisman_Levashov_Ovchinnikov_Sitnikov_Fortov_Khishchenko_2007}, the target was heated by a $p$-polarized laser pulse at the main laser wavelength $\lambda_1$ for the angle of incidence 45$^\circ$. The spatial distribution of the pumping radiation intensity on the target corresponded to the Gaussian with a focus beam diameter of about 70~$\mu$m at a level of $\exp(-2)$. A weak $s$-polarized probe pulse with varying time delay (the second harmonic $\lambda_2 = 620$~nm) was incident perpendicularly to the sample surface.

\section{Theoretical model}

The self-consistent theoretical model includes the system of electrodynamic equations for describing the absorption and reflection of laser radiation, ionization kinetic equations, and one-fluid hydrodynamic equations including electron-ion relaxation and electron heat conduction~\cite{ Andreev_Veysman_Efremov_Fortov_2003,Veysman_Cros_Andreev_Maynard_2006 }, as well as a new wide-range two-temperature equation of state of the irradiated substance. 

For the characteristic time intervals of processes under consideration ($t \leqslant 1$~ps), all the typical sizes of inhomogeneity in the $z$ direction perpendicular to the target surface do not exceed 1~$\mu$m. 
These are much smaller than the inhomogeneity sizes along the target surface, which are determined by the size of the focusing spot and are equal to tens of microns. 
Therefore, to analyze the experimental data~\cite{ Agranat_Andreev_Ashitkov_Veisman_Levashov_Ovchinnikov_Sitnikov_Fortov_Khishchenko_2007}, we use the one-dimensional version of the model developed that takes into account variations of all the quantities only in the $z$ direction and the single velocity component $V$ of the quasi-neutral motion of a plasma directed perpendicular to the target surface. 

The hydrodynamics equations of continuity for the volume concentration of heavy particles (atoms and ions) $n_a$ and changes in the substance momentum for the velocity $V$ as well as in the energies of electrons $e^e$ and heavy particles $e^i$ are written in the standard form including thermal ionization in the model of the average ion charge $Z$~\cite{ Andreev_Veysman_Efremov_Fortov_2003}, 
\begin{gather} 
\label{ngid} 
\partial_t n_{a} + \partial_z (n_{a} V) = 0, 
\\
\label{Ugid} 
[\partial_t + V \partial_z] V = - \rho^{-1} \partial_z (P_{e}+P_{i}) + F_p, 
\\
    n_a[\partial_t + V\partial_z]e^i = -P_i\partial_z V + Q^{ei},
    \label{hyd2}
\\
    Zn_a[\partial_t + V\partial_z]e^e = -\partial_z q_T - Q_Z +
    Q_{IB} - e^e\Theta - P_e\partial_z V - Q^{ei}, 
    \label{hyd1}
%
\\
\label{Z} 
[\partial_t + V \partial_z] Z = \Theta. 
\end{gather}
Here, $P_e$ and $P_i$ are the pressures of electrons and heavy particles respectively, determined by the equation of state of the substance; $\Theta$ and $Q_Z$ are the total rate of the thermal ionization and power density spent on ionization respectively, which are calculated using the model~\cite{ Andreev_Veysman_Efremov_Fortov_2003,Veysman_Cros_Andreev_Maynard_2006} of the average ion charge, the Lotz formula for impact ionization, and the detailed balance principle; $\rho = m_i n_a$ is the substance density, $m_i$ is the heavy particle mass, $Z = n_{e}/n_{a}$, $n_e$ is the concentration of electrons (quasi-neutrality is assumed), $F_p$ is the ponderomotive force, 
\begin{equation}
\label{F_p}
F_p = - K_1(\xi^\omega) \frac{Zm_e}{4m_{i}} \partial_z |\bs{V}_E|^2, 
\end{equation} 
$\bs{V}_E \equiv e |{\bs E}|(m_e\omega_1)^{-1}$ is the quiver velocity of electrons under the laser irradiation with the slowly varying in time amplitude of an electric field $\bs {E}$ and the frequency $\omega_1$, $e$ and $m_e$ are the electron charge and mass, $\xi^\omega = 3\sqrt{\pi}\nu_{\eff}(4\omega)^{-1}$, $\nu_{\eff}$ is the effective collision frequency of electrons, function 
\begin{equation}
\label{K_1}
K_{1}(x) = \frac{8}{3\sqrt{\pi}}\int\limits_0^\infty \frac {t ^ {10} \exp(-t^2)}{t^6 + x^2} \,\ud t 
\end{equation} 
arises due to integration of the velocity-dependent cross-section of electron-ion collisions over the Maxwell distribution function~\cite{Andreev_Veysman_Efremov_Fortov_2003}; $Q^{ei} = \gamma^{ei}Zn_a(T_e - T_i)$ is the electron-ion relaxation energy density, where $T_e$ and $T_i$ are the temperatures of electrons and heavy particles respectively, and the coefficient $\gamma^{ei}$ for lattice temperatures $T_i \lesssim T_\text{melt}$ ($T_\text{melt}$ is the melting temperature) is a constant ($\gamma^{ei} = 4.93 \cdot 10^{10}$~s$^{-1}$ for aluminum) and, for higher temperatures, is determined by the plasma formula $\gamma^{ei} = 3(m_e/m_i)\nu_\text{eff}$, where $m_e$ is the electron mass and $\nu_\text{eff}$ is the effective electron collision frequency; $q_T = K'T_e\partial_z T_e$ is the electron thermal flux, where the coefficient $K'$ for lattice temperatures $T_i \lesssim T_\text{melt}$ is a constant ($K' = 4.35 \cdot 10^{36}$~[erg~cm~s]$^{-1}$ for aluminum) and, for higher temperatures, is determined by the plasma formula~\cite{Spitzer_Harm_1953}, $K' = -128\kappa_Z Z n_a / (3\pi m_e \nu_\text{eff})$, the factor $k_Z \simeq 0.7$ represents the effect of electron-electron collisions on heat conduction; $Q_{IB}$ is the power density of the inverse bremsstrahlung absorption of the energy of the pump laser pulse, 
\begin{equation}
Q_{IB} = (8\pi)^{-1}\omega_1\text{Im}\,\{\varepsilon (\omega_1) \} |\bs{E}|^2, 
\end{equation}
where $\bs{E}$ is the electric-field amplitude of the laser pulse, $\varepsilon$ is the permittivity of matter, and $\omega_1$ is the heating (pump) radiation frequency. 

The slowly varying in time amplitude of the electric field strength $\bs{E} = E_x \bs{e}_{x} +E_z \bs{e}_{z}$ ($\bs{e}_{x}$ and $\bs{e}_{z}$ are the unit vectors, the plane $x0z$ is determined by the wave vector of incident radiation and the normal to the target surface, i.e. direction $0z$) for $p$-polarized pump laser pulse was expressed through the amplitude of the magnetic field $\bs {B} = B \bs{e}_{y}$ as $E_x = -i/(\varepsilon(\omega_1)k_1) \partial_z B$, $E_z = -B \sin(\theta)/\varepsilon(\omega_1)$, $k_1=\omega_1 / c$, $c$ is the velocity of light. The wave equation (with proper boundary conditions) for the laser magnetic field envelope $\bs {B}$~\cite {Ginzburg_1970}, 
\begin{equation}
\label{B}
\begin{array}{l}
\partial_z^2 B + k_1^2 [\varepsilon(\omega_1) - \sin^{2}(\theta)] B - \partial_z [\ln \varepsilon(\omega_1)] \partial_z B = 0,\\
\partial_z B|_{z = z_0} = i \varepsilon(\omega_1) k_1 \cos(\theta) [2E_{L} - B]|_{z = z_0},
\ B|_{z \to \infty} = 0,
\end{array}
\end{equation}
was solved numerically. Here $E_L = \sqrt{8\pi I_L / c}$, $I_L=I_L(t)$ is the intensity of incident laser pulse, $\theta$ is the angle of incidence (the angle between the direction $0z$ and the wave vector of incident radiation, $\theta=45^{\circ}$ for the discussed experiments), the point $z=z_0\leqslant 0$ is at the plasma--vacuum boundary position, and the laser pulse propagates from the left (before the laser heats up the target $z_0=0$, the target is at $z \geqslant 0$, i.e. at the right half-space).

Note that Eq.~(\ref{B}) describes also a collisionless resonance absorption, but in the discussed range of the pump laser intensities, $I \leqslant 2\cdot 10^{14}$~W/cm$^2$, for a high contrast laser pulse of 100~fs duration the characteristic scale length of the plasma density does not exceed a few nanometers and the electron plasma temperature is about 10~eV, and so the laser absorption takes place at over-critical densities in the normal skin heating regime~\cite{Gibbon_2005}.

The complex reflection coefficient of a weak $s$-polarized probe laser pulse with frequency $\omega_2$ at normal incidence was determined in the linear approximation. For this goal the wave equation~\cite{Ginzburg_1970} with boundary conditions for single component of electric field of the probe laser pulse $\bs{E} = E(z) \bs{e}_{y}$, 
\begin{equation}
\label{E}
\begin{array}{l}
\partial^{2}_z E + k_2^2 [\varepsilon(\omega_2) - \sin^{2}(\theta)] E = 0,
\
k_2=\omega_2 / c,\\
\partial_z E|_{z = z_0} = i k_2 \cos(\theta) [2E_{L} - E]|_{z = z_0},
\ E_{z \to \infty} = 0,
\end{array}
\end{equation}
was solved numerically with $\theta = 0^{\circ}$ on the spatially non-uniform profiles of permittivity $\varepsilon(\omega_2)$ for different time delays. For each time delay the permittivity is determined as described below with the help of numerical solution of Eqs.~(\ref{ngid})--(\ref{Z}) that describe the hydrodynamics of a target irradiated by the pump laser pulse determined self-consistently by Eq.~(\ref{B}).

It should be noted, that in Eqs.~(\ref{B}) and~(\ref{E}) for subpicosecond laser pulses the transverse (to the normal of the target, $0z$) gradients of fields and plasma permittivity, which are determined by characteristic scale lengths of the order of ten micrometers, can always be omitted in comparison with longitudinal gradients, which are determined by the skin-layer depth of a typical size of order of ten nanometers. 

The effective collision frequency $\nu_\text{eff}$ over the entire temperature range is determined as the minimum of the three values, 
\begin{equation}
\nu_\text{eff} = \text{min}\{\nu_\text{met}, \nu_\text{pl}, \nu_\text{max}\}, 
\end{equation}
where $\nu_\text{met}$ is the effective collision frequency in the metal plasma for $T_e \lesssim T_{\Fermi} = (3\pi^2Zn_a){}^{2/3}\hbar^2/(2m_e)$, $\nu_\text{pl}$ is the collision frequency for the weakly coupled plasma~\cite{Silin_Rukhadze_1961}, 
\begin{equation}
\nu_\text{pl} = (4/3)\sqrt{2\pi} Z^2 n_a e^4 \Lambda / \sqrt{m_e T_e\!^3},
\end{equation}
$\Lambda$ is the Coulomb logarithm, and $\nu_\text{max}$ is the maximum collision frequency determined by the condition~\cite{Yakubov_1993} that the collision mean free path of electrons $\lambda_{\text{p}e} \sim v_e / \nu_\text{eff}$ ($v_e$ is the mean electron velocity) is no less than the mean distance between ions $r_0 \sim n_a\!^{-1/3}$, 
\begin{equation}
\label{numax}
    \nu_\text{max} = \C_1\omega_{\text{p}e},
\end{equation}
where $\omega_{\text{p}e} = \sqrt{4\pi Z n_a e^2 / m_e}$ is the electron plasma frequency and the numerical coefficient $\C_1 \lesssim 1$ is chosen by fitting the calculations to the experimental data~\cite{ Agranat_Andreev_Ashitkov_Veisman_Levashov_Ovchinnikov_Sitnikov_Fortov_Khishchenko_2007}. Despite the simplicity of such an approach to the determination of $\nu_\text{eff}$, it ensures satisfactory accuracy as compared to much more complicated methods~\cite{Semkat_Redmer_Bornath_2006}. 

The effective collision frequency in the metal
plasma $\nu_\text{met}$ is determined as
\begin{gather}
\label{numet}
\nu_\text{met} = C_{e \cdot ph}T_i / \hbar + \C_2 T_e\!^2 / \hbar T_{\Fermi},
%
\\
\label{Ceph}
C_{e \cdot ph} = C_{00} + C_{0}[1 - \min\{T_i / T_\text{melt},1\}^{1/2}]. 
\end{gather}
The first and second terms in Eq.~(\ref{numet}) represent the contributions from the electron-phonon~\cite{Fisher_Fraenkel_Henis_Moshe_Eliezer_2001} and electron-electron~\cite{Abrikosov_1988} collisions respectively. The constant $\C_2$ is chosen by comparing with the experimental data (also see Ref.~\citenum{Isakov_Kanavin_Uryupin_2006_10}). The first and second terms in Eq.~(\ref{Ceph}) represent the contributions from the intraband and interband transitions~\cite{ Fisher_Fraenkel_Henis_Moshe_Eliezer_2001,Ashcroft_Mermin_1976} respectively. The constant $C_{00}$ is determined from the data on the static conductivity of the metals~\cite{Ashcroft_Mermin_1976} (for aluminum, $C_{00} \simeq 3.28$). The constant $C_0$ is chosen so as to ensure the tabulated reflection coefficient $|r|^2$ of the metal under consideration at room temperature. For aluminum, $|r|^2 \simeq 0.96$ and 0.91, whereas $C_0 \simeq 23$ and 10 for the heating and probe pulses wavelengths $\lambda_1 = 1.24$~$\mu$m and $\lambda_2 = 0.62$~$\mu$m respectively. When the lattice temperature $T_i$ exceeds the melting temperature $T_\text{melt}$, the band structure of the metal is destroyed and the contribution from the interband transitions to the electron-phonon collision frequency vanishes~\cite{Palik_1985}. This circumstance is taken into account by a phenomenological dependence on $T_i / T_\text{melt}$ in Eq.~(\ref{Ceph}).

The permittivity of substance is determined by the Drude formula for the metallic plasma~\cite{Ashcroft_Mermin_1976} for $T_e \leqslant T_1 = 0.75T_{\Fermi}$ and by the formula for a weakly coupled non-degenerate plasma~\cite{Silin_Rukhadze_1961, Andreev_Veysman_Efremov_Fortov_2003, Veysman_Cros_Andreev_Maynard_2006} for $T_e \geqslant T_2 = 1.5T_{\Fermi}$. In the interval $T_1 < T_e < T_2$, a linear interpolation between the Drude formula and plasma formula is used with the above-indicated $\nu_\text{pl}$ value. The optical electron mass in the Drude formula is taken as $m_\text{opt} = 1.5m_e$~\cite{Palik_1985}. 

The thermodynamic characteristics of the condensed phase of the target substance for both the thermal equilibrium between ions and electrons and non-equilibrium heating (when $T_e > T_i$) are determined by using a new semiempirical equation of state in a wide region of densities and temperatures. In this equation of state, the free energy $F( \rho, T_i, T_e, Z)$ is represented as the sum of two terms, $F = F_i(\rho, T_i) + F_e(\rho, T_e, Z)$, determining the contributions of heavy particles and electrons respectively. 

The first term, $F_i = F_c(\rho) + F_a(\rho, T_i)$, includes the energy of the interaction between heavy particles and electrons at $T_i = T_e = 0$ ($F_c$) and the contribution from the thermal motion of heavy particles ($F_a$). The dependence of cold energy $F_c(\rho)$ is determined by the procedure described in Ref.~\citenum{Khishchenko_2004}, which ensures the equality of the total pressure in the system to atmospheric pressure at normal density (for aluminum, $\rho_0 = 2.71$~g/cm$^3$) and room temperature, as well as the agreement with the available data of the shock-wave experiments and Thomas--Fermi calculations with quantum and exchange corrections for the high energy densities. The thermal contribution of heavy particles to the free energy is given by the expression~\cite{Bushman_Lomonosov_Fortov_1992_metals_eng} 
\begin{equation}
    F_a(\rho, T_i) = \frac{3T_i}{2m_i}\ln 
    \bigg( \frac{\Theta_a\!\!^2}{T_i\!^2} + \frac{T_a\sigma^{2/3}}{T_i} \bigg), 
\end{equation}
where $\sigma = \rho/ \rho_0$. To determine the dependence of the characteristic temperature $\Theta_a = \Theta_a(\rho)$, we use the interpolation formula~\cite{Altshuler_Bushman_Zhernokletov_Zubarev_Leontev_Fortov_1980} 
\begin{equation}
    \Theta_a(\rho) = \sigma^{2/3} \exp \bigg[
        (\gamma_{0a} - 2/3)\frac{B_a\!\!^2 + D_a\!\!^2}{B_a} \arctan\bigg(
            \frac{B_a \ln\sigma}{B_a\!\!^2 + D_a(\ln\sigma + D_a)}
        \bigg)
    \bigg],
\end{equation}
where $\gamma_{0a}$ is the Gr\"uneisen coefficient under normal conditions. The constants $T_a$, $B_a$, and $D_a$ are determined from the requirement of the optimal description of the experimental data on the thermal expansion and shock compressibility of porous samples of the substance. 

The free energy of the electron gas in the metal is given by the expression 
\begin{equation}
\label{Fe}
    F_e(\rho, T_e, Z) = -\frac{3Z T_e }{2m_i} \ln \bigg(
        1 + \frac{\pi^2 T_e}{6T_{\Fermi}}
    \bigg).
\end{equation}
Equation~(\ref{Fe}) for low and high temperatures is an equation for an ideal degenerate Fermi gas and an ideal Boltzmann gas of free electrons respectively~\cite{Landau_Lifshitz_V_1980}. Expressions similar to Eq.~(\ref{Fe}) were previously used in the equations of state in Refs.~\citenum{Altshuler_Bushman_Zhernokletov_Zubarev_Leontev_Fortov_1980} and~\citenum{Basko_1985}. The internal energies of one heavy particle and electron, as well as the total pressures for heavy particles and electrons, are expressed in terms of $F$ as $e^i = m_i[F_i - T_i(\partial F_i / \partial T_i)_\rho]$ and $e^e = m_iZ^{-1}[F_e - T_e(\partial F_e / \partial T_e)_{\rho, Z}]$, as well as $P_i = \rho^2(\partial F_i / \partial\rho)_{T_i}$ and $P_e = \rho^2(\partial F_e / \partial\rho)_{T_e, Z}$ respectively.

\section{Simulation results}

\begin{figure}[!ht]
\centering{
\includegraphics[width=0.7\columnwidth]{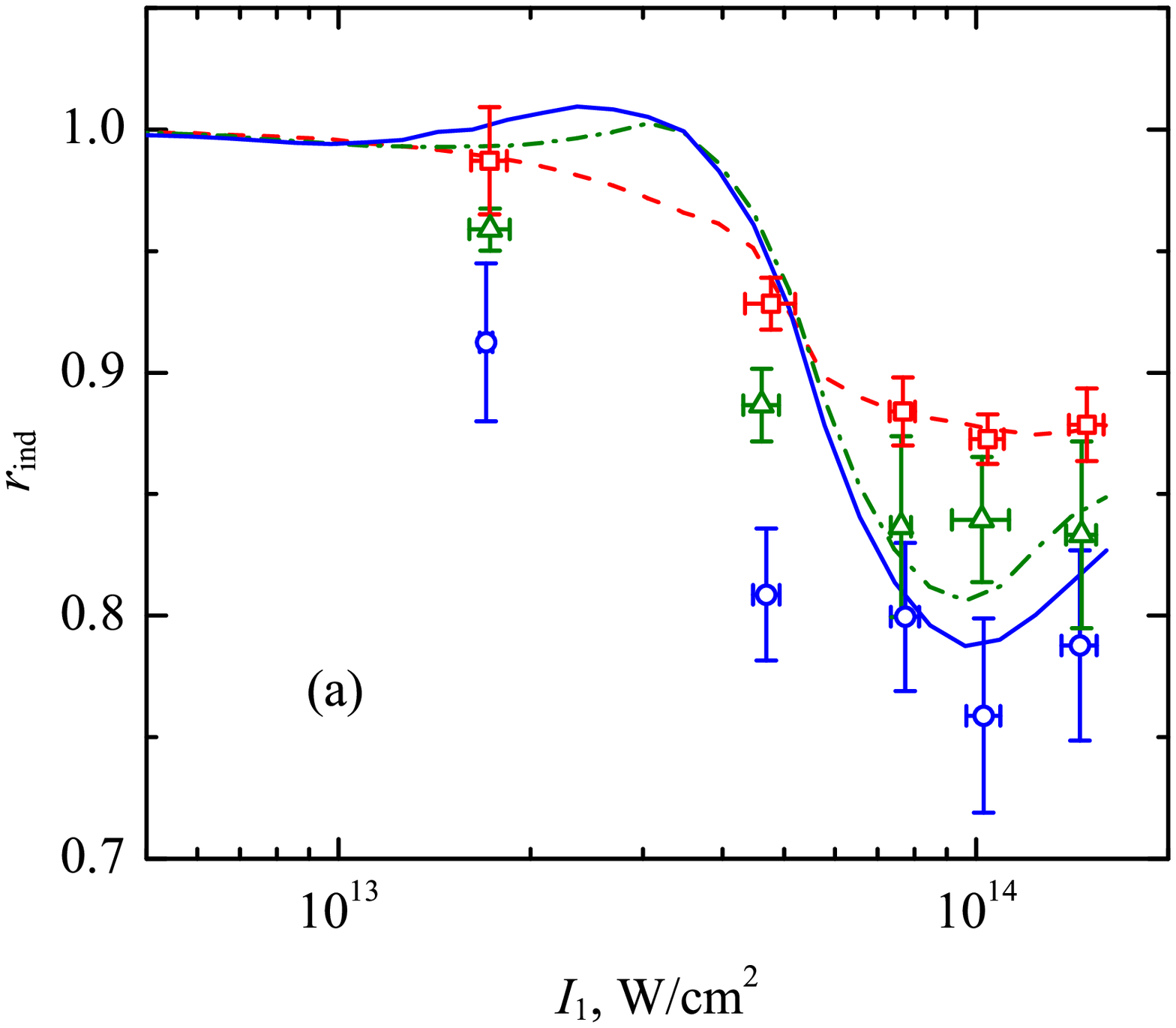}
\includegraphics[width=0.7\columnwidth]{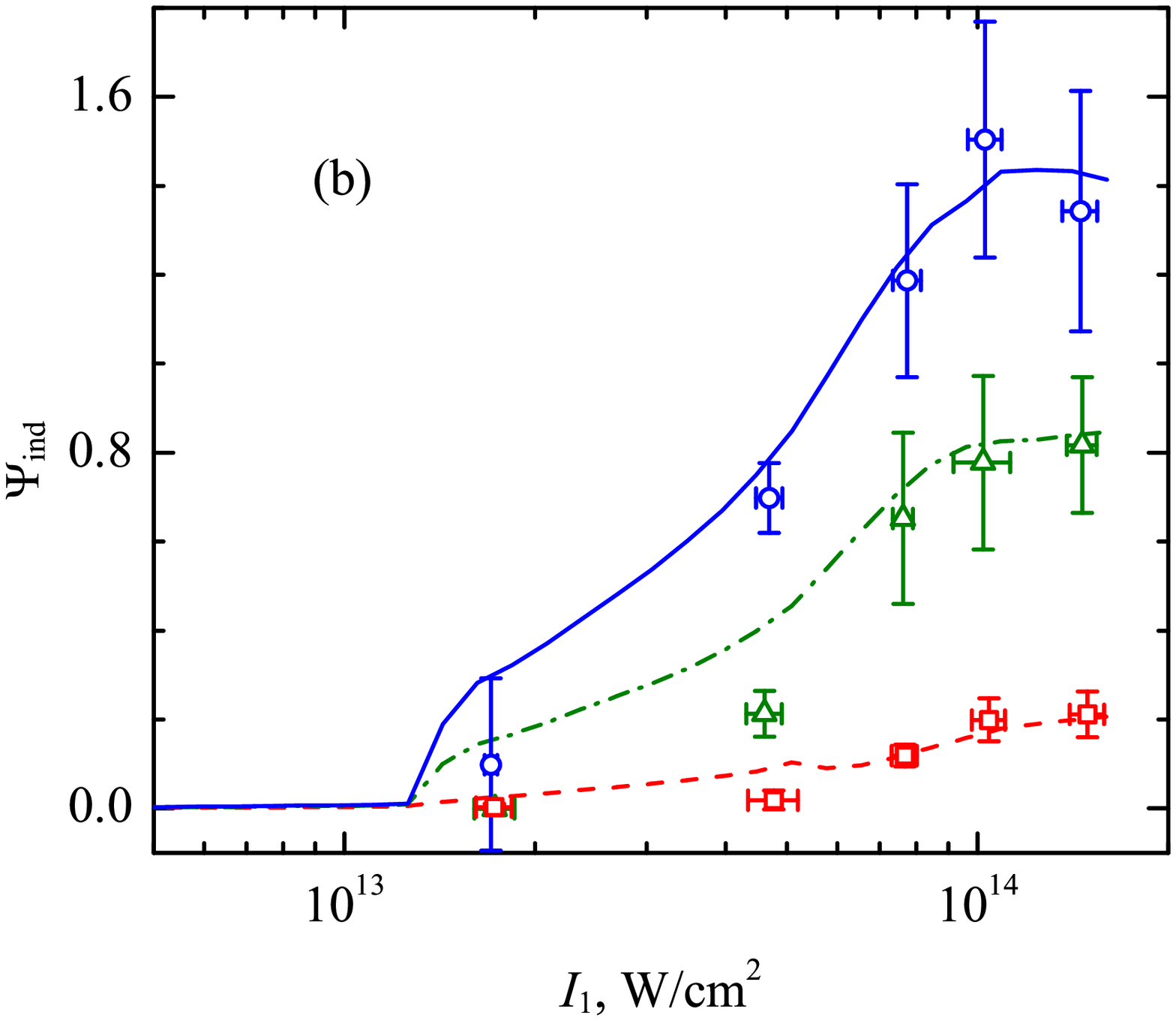}
}
\caption{
Experimental points~\cite{ Agranat_Andreev_Ashitkov_Veisman_Levashov_Ovchinnikov_Sitnikov_Fortov_Khishchenko_2007} and calculated curves for (a) $r_\text{ind}$ and (b) $\Psi_\text{ind}$ versus the intensity $I_1$ of the pump pulse for time delays (squares and dashed lines) $\Delta t = 130$, (triangles and dash-dot lines) 530, and (circles and solid lines) 930~fs. 
}
\label{f:rI}
\end{figure}

Figure~\ref{f:rI} shows the results of modeling as well as the experimental data~\cite{ Agranat_Andreev_Ashitkov_Veisman_Levashov_Ovchinnikov_Sitnikov_Fortov_Khishchenko_2007 } on the absolute value ($r_\text{ind}$) and the phase ($\Psi_\text{ind}$) of the complex reflection coefficient of aluminum as functions of the maximum intensity $I_1$ of a heating laser pulse for various time delays ($\Delta t = 130$, 530, and 930~fs) of the pump and probe pulses. The theoretical lines in Fig.~\ref{f:rI} are obtained by simulating the heating of the target by the pump laser pulse, and also by the calculation of the amplitude and phase of the reflected probe pulse at the doubled frequency using described above theoretical model. 

Comparison of the experimental data with the calculations of the absolute value $r_\text{ind}$ and phase $\Psi_\text{ind}$ of the complex reflection coefficient of the probe pulse (see Fig.~\ref{f:rI}) makes it possible to determine (in the framework of the present model) the important properties of the strongly coupled solid-density plasma such as the maximum effective frequency of electron momentum relaxation given by Eq.~(\ref{numax}) and the contribution to the effective electron-electron collision frequency specified by Eq.~(\ref{numet}). The best agreement with the experimental data for the aluminum targets is reached at $\C_1 \simeq 0.3$, which is close to the theoretical estimate~\cite{Yakubov_1993}, and at $\C_2 \simeq 0.85$ in Eqs.~(\ref{numax}) and~(\ref{numet}) for the effective collision frequencies. In this case, the uncertainty in the choice of the coefficients does not exceed 15\% including the experimental errors. 

For the indicated parameters, the proposed model well reproduces $r_\text{ind}$ and the phase $\Psi_\text{ind}$ as functions of the laser pumping pulse intensity $I_1$ for all time delays $\Delta t$ between the pump and probe pulses in the measurements with $I_1 \gtrsim 5 \cdot 10^{13}$~W/cm$^2$. For fluxes $I_1 \lesssim 5 \cdot 10^{13}$~W/cm$^2$, the model is in agreement with the experimental data for $\Delta t < 500$~fs, whereas significant discrepancies exist for $r_\text{ind}$ at $\Delta t > 500$~fs. These discrepancies can be caused by the formation of a two-component mixture consisting of the low-density plasma and condensed-phase fragments on the target surface~\cite{Povarnitsyn_Itina_Sentis_Khishchenko_Levashov_2007}. A detailed description of this effect is beyond the scope of the model and is the subject of further investigations. 

\begin{figure}[!ht]
\centering{
\includegraphics[width=0.7\columnwidth]{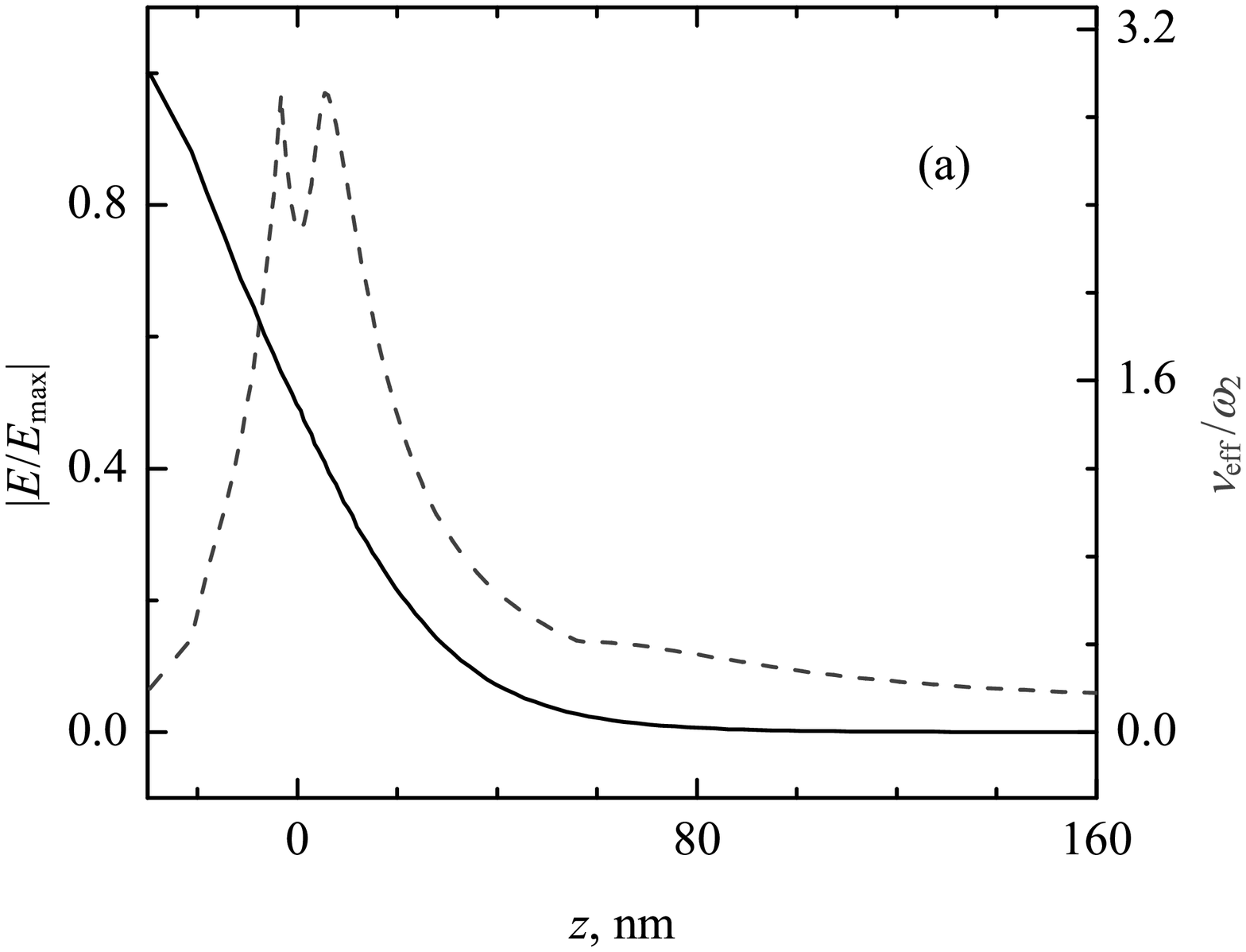}
\includegraphics[width=0.7\columnwidth]{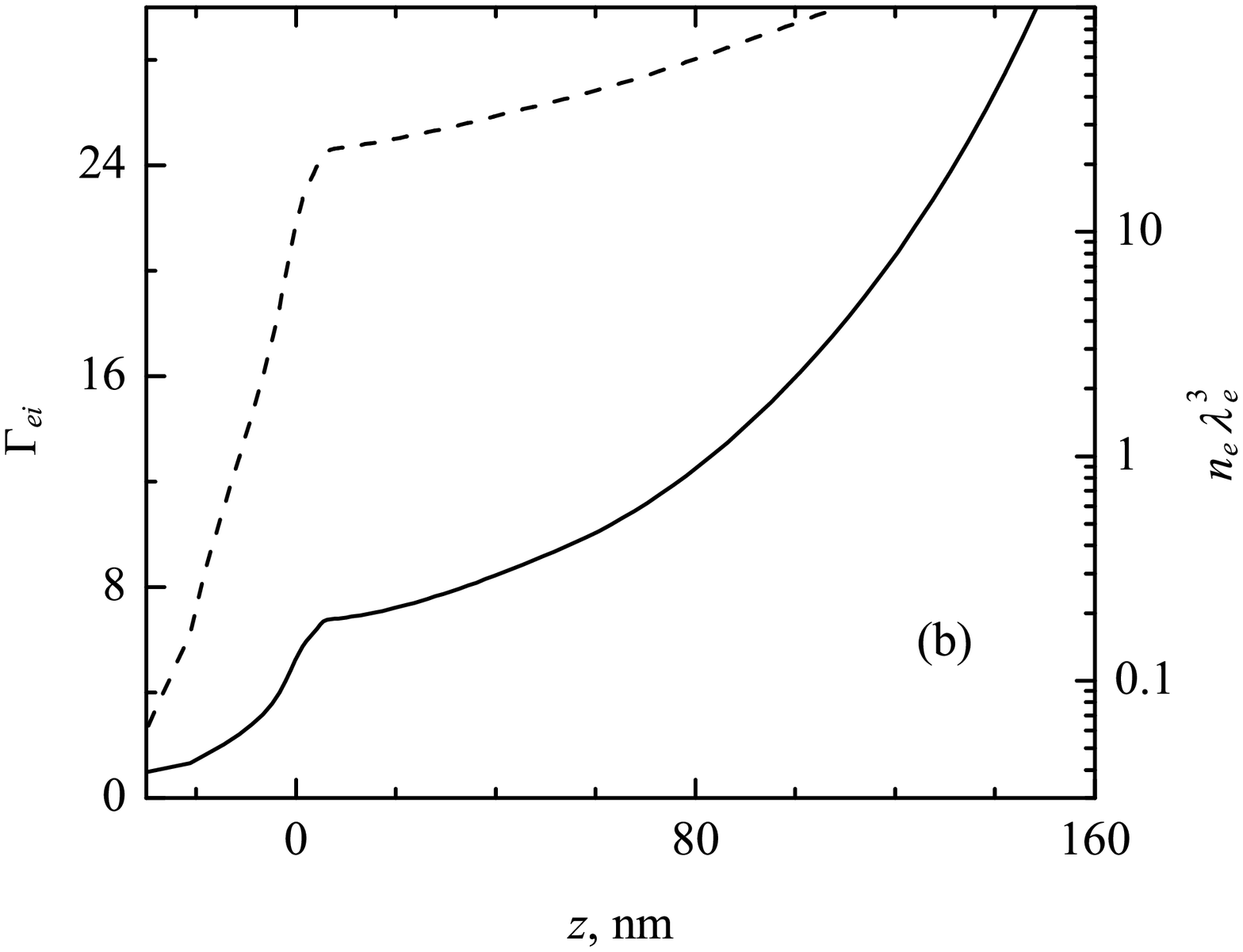}
}
\caption{
Target-depth dependencies calculated for the aluminum plasma characteristics at $\Delta t = 530$~fs and $I_1 = 6.7 \cdot 10^{13}$~W/cm$^2$: 
(a) normalized (solid line) electric field strength of the probe pulse and (dashed line) effective frequency of electron collisions, (b) the parameters of (solid line) non-ideality and (dashed line) degeneration. 
}
\label{f:Hx}
\end{figure}

Figure~\ref{f:Hx} shows the effective collision frequency and the field strength of the laser probe pulse, as well as the non-ideality parameter $\Gamma_{ei} = Ze^2 / (n_a\!^{-1/3} T_e)$ and the degeneration parameter $n_e \lambda_e\!\!^3 = (8\pi/3)(T_{\Fermi}/T_e){}^{3/2}$ as functions of the target depth for the fixed intensity $I_1 = 6.7 \cdot 10^{13}$~W/cm$^2$ and the time delay $\Delta t = 530$~fs. Note that the average ion charge for the parameters under consideration does not change in the calculation time and remains equal to its initial value $Z = 3$. 

As is seen in Fig.~\ref{f:Hx}, the plasma parameters are strongly inhomogeneous in the skin-layer region, where the reflected-signal field of the laser probe pulse is formed. For this reason, despite short duration of the processes under consideration, the Fresnel formulas describing the reflection from a \textit{homogeneous} medium with a stepwise boundary are inapplicable to the calculation of the reflection coefficient and its phase. 

Analysis of the simulation results shown in Fig.~\ref{f:Hx} indicates that the plasma formed on the target surface for the experimental parameters under consideration is in a strongly coupled state ($\Gamma_{ei} \gtrsim 1$) and is strongly degenerate ($n_e \lambda_e\!\!^3 \gtrsim 1$) in the entire region except for the plasma corona. 

\begin{figure}[!ht]
\centering{
\includegraphics[width=0.7\columnwidth]{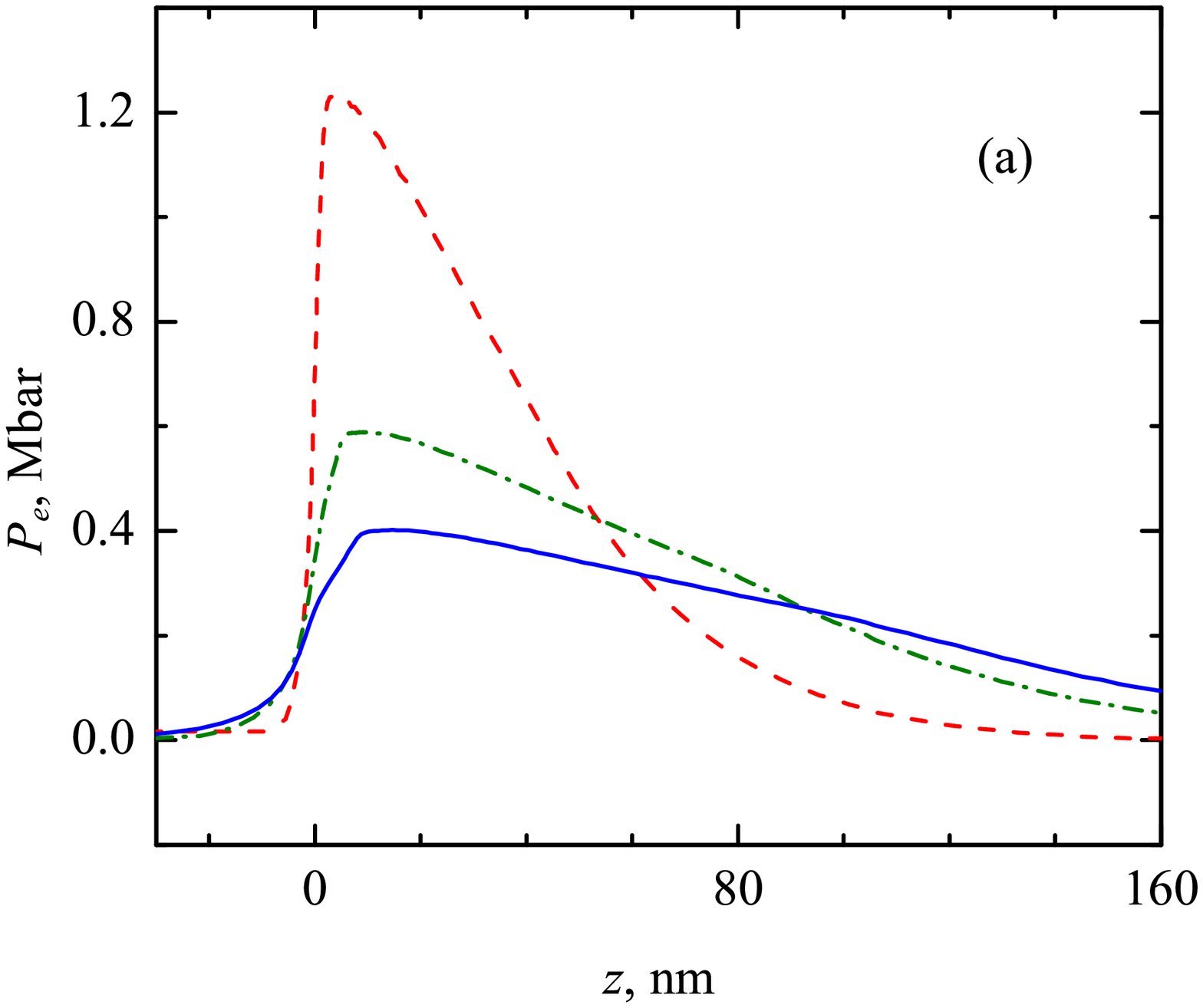}
\includegraphics[width=0.7\columnwidth]{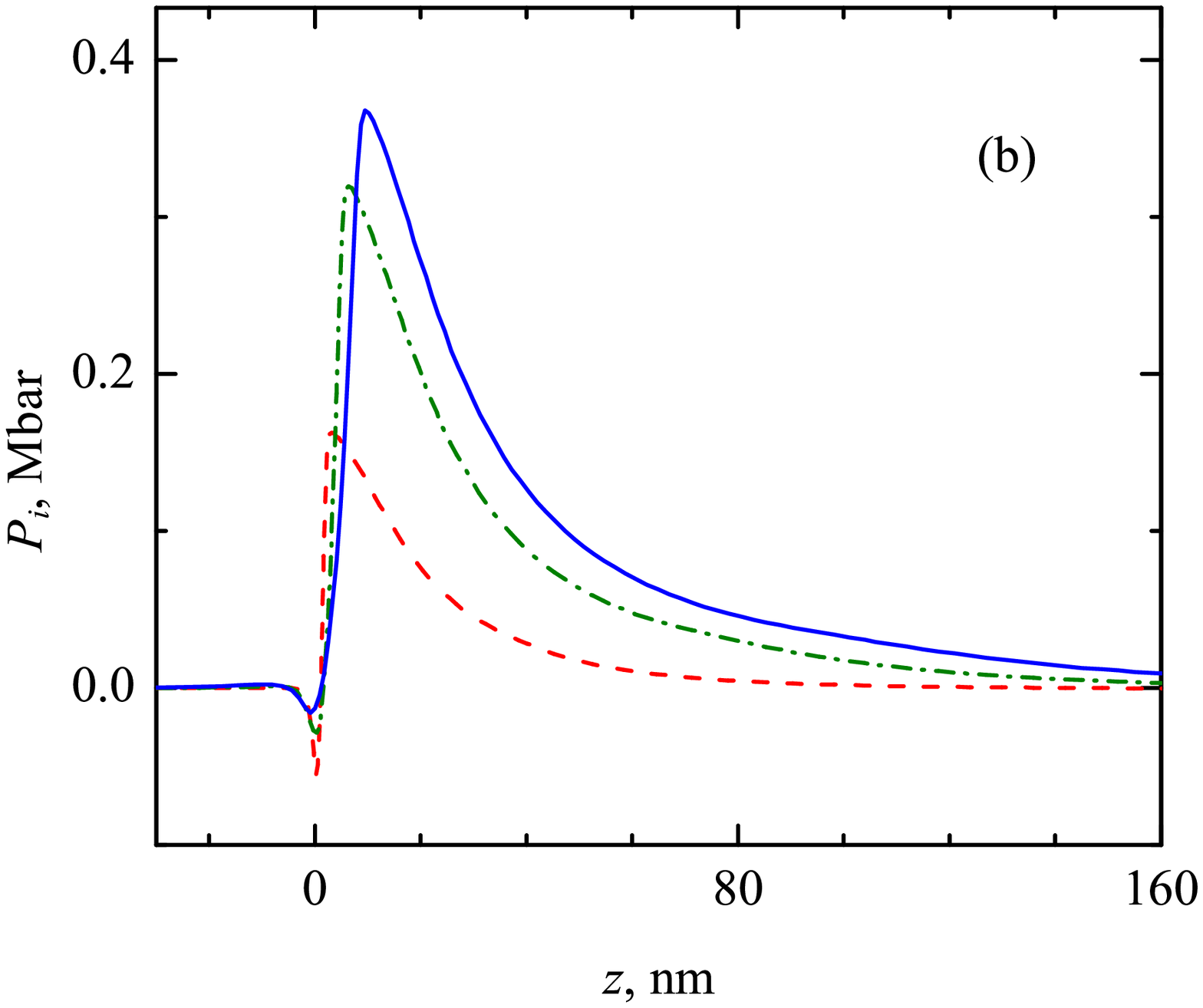}
}
\caption{
Calculated target-depth distribution of the (a) electron and (b) heavy particle pressures in the aluminum plasma at $I_1 = 6.7 \cdot 10^{13}$~W/cm$^2$ and (dashed lines) $\Delta t = 130$, (dash-dot lines) 530, and (solid lines) 930~fs. 
}
\label{fig3}
\end{figure}

Figure~\ref{fig3} illustrates the formation of a shock wave propagated into the target, which is irradiated by the pump laser pulse of $I_1 = 6.7 \cdot 10^{13}$~W/cm$^2$ at different time delays. For the times $\Delta t \lesssim 1$~ps, both the pressure and temperature of electrons are higher than those of heavy particles are. The heavy particle velocity at $\Delta t = 930$~fs in the shock wave is of $V \simeq 0.3$~km/s. In the direction opposite to the shock-wave propagation, the plasma is expanded with the particle velocity of $V \sim 10$~km/s. Under the rarefaction, a region of negative pressures takes place, see Fig.~\ref{fig3}(b). Such behavior of pressure causes a fragmentation of irradiated target at later stages of the considering process~\cite{ Povarnitsyn_Itina_Sentis_Khishchenko_Levashov_2007 }.

\section{Conclusion}

Thus we proposed semiempirical models of optical, transport, and thermodynamic properties of Al over a wide range of temperatures and pressures. Those allowed us to have carried out the two-temperature hydrodynamic simulation of the processes in the metal under the action of intense femtosecond laser pulses. We emphasize that the data on the amplitude and the phase of the reflected field of the probe laser pulse from the experiment~\cite{ Agranat_Andreev_Ashitkov_Veisman_Levashov_Ovchinnikov_Sitnikov_Fortov_Khishchenko_2007 } make it possible to acquire important information on the transport properties of the strongly coupled plasma formed under the action of the pump laser radiation on the target surface in the subpicosecond time intervals. It is shown that the plasma inhomogeneity decisively affects the reflective properties of the target even in such short times.

\acknowledgments

This work was supported in part by the Russian Foundation for Basic Research (projects No.~07-02-92160 and 08-08-01055) and the Council of the President of the Russian Federation for Support of Young Russian Scientists and Leading Scientific Schools (project No.~NSh-6494.2008.2).


\hyphenation{Post-Script Sprin-ger}

\newpage 

\newpage 

\newpage 

\end{document}